# Dirac nodal line metal for topological antiferromagnetic spintronics

Ding-Fu Shao,[1,*] Gautam Gurung,[1] Shu-Hui Zhang,[2] and Evgeny Y. Tsymbal[1,†]

[1] *Department of Physics and Astronomy & Nebraska Center for Materials and Nanoscience,*
*University of Nebraska, Lincoln, Nebraska 68588-0299, USA*

[2] *College of Science, Beijing University of Chemical Technology,*
*Beijing 100029, People's Republic of China*

Topological antiferromagnetic (AFM) spintronics is an emerging field of research, which exploits the Néel vector to control the topological electronic states and the associated spin-dependent transport properties. A recently discovered Néel spin-orbit torque has been proposed to electrically manipulate Dirac band crossings in antiferromagnets; however, a reliable AFM material to realize these properties in practice is missing. In this letter, we predict that room temperature AFM metal $MnPd_2$ allows the electrical control of the Dirac nodal line by the Néel spin-orbit torque. Based on first-principles density functional theory calculations, we show that reorientation of the Néel vector leads to switching between the symmetry-protected degenerate state and the gapped state associated with the dispersive Dirac nodal line at the Fermi energy. The calculated spin Hall conductivity strongly depends on the Néel vector orientation and can be used to experimentally detect the predicted effect using a proposed spin-orbit torque device. Our results indicate that AFM Dirac nodal line metal $MnPd_2$ represents a promising material for topological AFM spintronics.

The discovery of novel quantum phenomena in solids, resulting from the interplay between the electron, spin, and orbital degrees of freedom, enriches a continuously evolving field of spintronics and opens opportunities to enhance the efficiency of electronic devices [1]. Recently, antiferromagnetic (AFM) spintronics has emerged as a subfield of spintronics, where an AFM order parameter also known as the Néel vector is exploited to control spin-dependent transport properties [2-4]. Due to being robust against magnetic perturbations, producing no stray fields, and exhibiting ultrafast dynamics, antiferromagnets can serve as promising functional materials for spintronic applications, which may expand to very diverse areas ranging from terahertz information technologies to artificial neural networks [5].

The interest to AFM spintronics has largely been stimulated by the recent discovery of electrical switching of a collinear antiferromagnet by spin-orbit torque [6]. It is known that spin-orbit torques can originate from the inverse spin-galvanic effect [7], which occurs in magnetic materials with broken space-inversion symmetry due to spin-orbit coupling (SOC) [8-11]. If an antiferromagnet is formed of two antiparallel-aligned spin sublattices, whose atomic structure has broken space-inversion symmetry but the sublattices form space-inversion partners, the inverse spin-galvanic effect produces a nonequilibrium local spin polarization of opposite sign on the two spin sublattices. The resulting staggered effective magnetic field generates an alternating in sign spin-orbit torque, known as the Néel spin-orbit torque, on the sublattice magnetizations, thus acting with a torque on the Néel vector [12,13]. The control of the Néel vector by electric current has been realized using tetragonal CuMnAs [6] and $Mn_2Au$ [14] antiferromagnets, thus demonstrating a viable approach to the AFM-based memories [15] and providing a route to ultra-fast spintronic devices [4,5].

In parallel with these developments, there has been increasing interest in materials and structures where quantum effects are responsible for novel physical properties, revealing the important roles of symmetry, topology, and dimensionality [16]. Among such quantum materials are graphene [17], topological insulators [18], Dirac and Weyl semimetals [19], and beyond [20]. These materials are characterized by non-trivial fermionic excitations resulting from discrete band crossings as well as continuous degenerate states, such as the nodal lines [21,22] and their exotic connections [23,24]. Using the unique properties of the novel fermionic states has been envisioned for spintronics applications [25,26]. A particular example is the demonstration of significantly enhanced spin-orbit torques in ferromagnet/topological insulator heterostructures [27,28].

The discovery of the electrical control of the Néel vector [6,14] opens a new direction in spintronics, involving the interplay between the topological electronic states and antiferromagnetism [29, 30]. A notable example is the proposed control of Dirac quasiparticles in an antiferromagnet by reorientation of the Néel vector [31]. ON and OFF switching of the symmetry protection of the Dirac band crossing has been predicted in an AFM Dirac semimetal, such as orthorhombic CuMnAs [32], resulting in the topological metal-insulator transition (Figs. 1(a-d)) and topological anisotropic magnetoresistance [31].

To realize these properties in practice, however, the Dirac quasiparticles are required to appear precisely at the Fermi energy ($E_F$), demanding a strict control of the stoichiometry and structural quality of the sample, which is not easy to achieve in real experimental conditions. On the other hand, there exist a class of quantum materials exhibiting Dirac nodal lines within a broad energy window including those crossing the Fermi energy. In an AFM material with such a nodal line, the Néel spin-orbit torque control of the Dirac band crossings would be much easier to realize and detect in transport measurements (Figs. 1(e,f)).

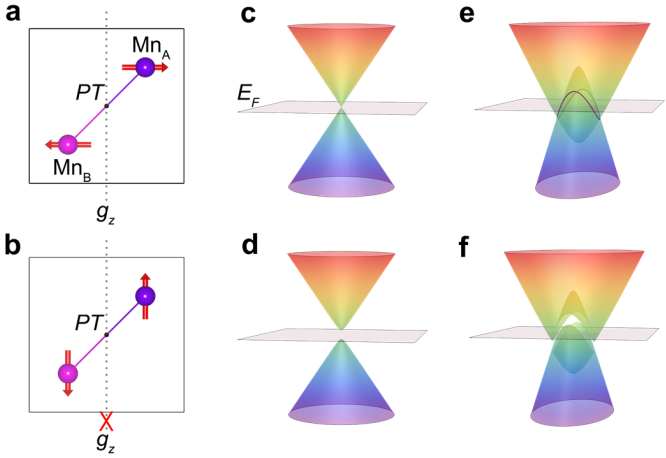

FIG. 1. Controlling a Dirac point or a Dirac nodal line by the Néel vector. (a,b) Schematics of an antiferromagnet with two magnetic sublattices (denoted as $Mn_A$ and $Mn_B$) connected by the $PT$ symmetry for two orientations of the Néel vector, preserving (a) and breaking (b) glide symmetry $g_z$ (indicated by the dotted lines). Red arrows indicate the magnetic moments. (c-f) Schematics of the band structure around the Fermi energy ($E_F$) for a Dirac point (c, d) or a dispersive Dirac nodal line (e, f) for preserved (c, e) or broken (d, f) glide symmetry $g_z$.

In this letter, we predict that AFM metal $MnPd_2$ has the desired properties: it has the required symmetry to support the Néel spin-orbit torque and holds a dispersive Dirac nodal line across the Fermi energy. The reorientation of the Néel vector leads to switching between the symmetry protected degenerate state and the gapped state associated with the Dirac nodal line. We show that the spin Hall conductivity of $MnPd_2$ strongly depends on the Néel vector orientation and can be used to detect the effect. $MnPd_2$ has been synthesized in the laboratory, has the Néel temperature ($T_N$) well above the room temperature, and thus represents a new promising material for topological antiferromagnetic spintronics.

Figure 2(a) shows the crystal structure of orthorhombic $MnPd_2$ which belongs to nonsymmorphic space group $Pnma$ [33]. In the paramagnetic phase, $MnPd_2$ has time-reversal symmetry $T$, space-inversion symmetry $P$, three glide planes, and three screw axes. Neutron diffraction reveals a collinear AFM ordering up to $T_N = 415 \pm 10$ K. In the ground state, the Néel vector lies along the [010] direction with the magnetic moments of the Mn atoms being parallel in the (010) planes but antiparallel between the successive (010) planes [33]. As seen from Fig. 2(a), the inversion-partner sites are occupied by the Mn atoms with oppositely oriented magnetic moments. Such an AFM ordering breaks both the $P$ and $T$ symmetries but the combined $PT$ symmetry is preserved. This condition is sufficient to produce the Néel spin-orbit torque on the Néel vector by passing an electric current [31]. Our calculation of the total energy predicts the lower energy for the Néel vector ($\vec{n}$) lying in the (100) plane with the magnetocrystalline anisotropy energy of $E_{[001]} - E_{[010]} = 0.14$ meV/f.u. This result is consistent with the experiment [33], showing that the easy axis lies along the [010] direction ($\vec{n} \parallel [010]$). The calculated magnetic moment of 3.89 $\mu_B$ per Mn atom is also in agreement with that (4.0 $\mu_B$) found by this experiment.

The AFM ordering determines the magnetic space group symmetry of $MnPd_2$, as shown in Table S1 in the Supplemental Material [34]. It is evident that the nonsymmorphic symmetries can be turned on and off by rotating the Néel vector. As we will see below, this leads to the ability to close and open a gap at the Dirac nodal line and thus to control spin-dependent transport properties of $MnPd_2$.

Our density functional theory (DFT) calculations [34] predict that around the Fermi energy, the electronic band structure of $MnPd_2$ is represented by the Mn and Pd $d$-orbitals (Fig. S1). Due to the preserved $PT$ symmetry, every band is doubly degenerate. As seen from Fig. S2, there are three bands crossing $E_F$. Below we focus on the bands around $E_F$ lying in the $k_z = \frac{\pi}{c}$ plane, which form a Dirac nodal line.

Figures 2(c,d) show the calculated band structure along the high symmetry paths in the $k_z = \frac{\pi}{c}$ plane for $\vec{n} \parallel [001]$ and $\vec{n} \parallel [010]$, respectively. For $\vec{n} \parallel [001]$, there are three four-fold degenerate crossings points: at $E = -0.071$ eV along the Z-U line, at $E = 0.234$ eV along the T-Z line, and at $E = 0.192$ eV along the Z-R line (indicated by arrows in Fig. 2(c)). These crossings are protected by the glide symmetry $g_z = \{M_z | (\frac{1}{2}, 0, \frac{1}{2})\}$ [34], leading to a loop-like Dirac nodal line in the $k_z = \frac{\pi}{c}$ plane surrounding the Z point (Fig. 2(e)). This Dirac nodal line is dispersive covering a wide energy window ranging from $E \approx -0.07$ eV to $E \approx 0.45$ eV.

Reorientation of the Néel vector breaks the $g_z$ symmetry (Table S1) and opens an energy gap along the Dirac nodal line, as seen from Figs. 2(d,f) for $\vec{n} \parallel [010]$. Although the gaps at the crossing points along the Z-U and T-Z directions are relatively small (a few meV), along the Z-R direction the gap exceeds 30 meV (Fig. 2(d)). We find that along the Dirac nodal line shown in Fig. 2(f), the gap varies from about 1 meV to 45 meV and is about 20 meV at the Fermi energy. Thus, we conclude that $MnPd_2$ exhibits the sought material properties: it has the required symmetry to support the Néel spin-orbit torque and holds a dispersive Dirac nodal line across the Fermi energy, which is gapped by the reorientation of the Néel vector.

Along with the electronic structure of $MnPd_2$, the Néel vector controls its spin-dependent transport properties, such as the spin Hall effect [35]. Below we calculate the spin Hall conductivity of $MnPd_2$, and show that it is strongly affected to the orientation of the Néel vector.

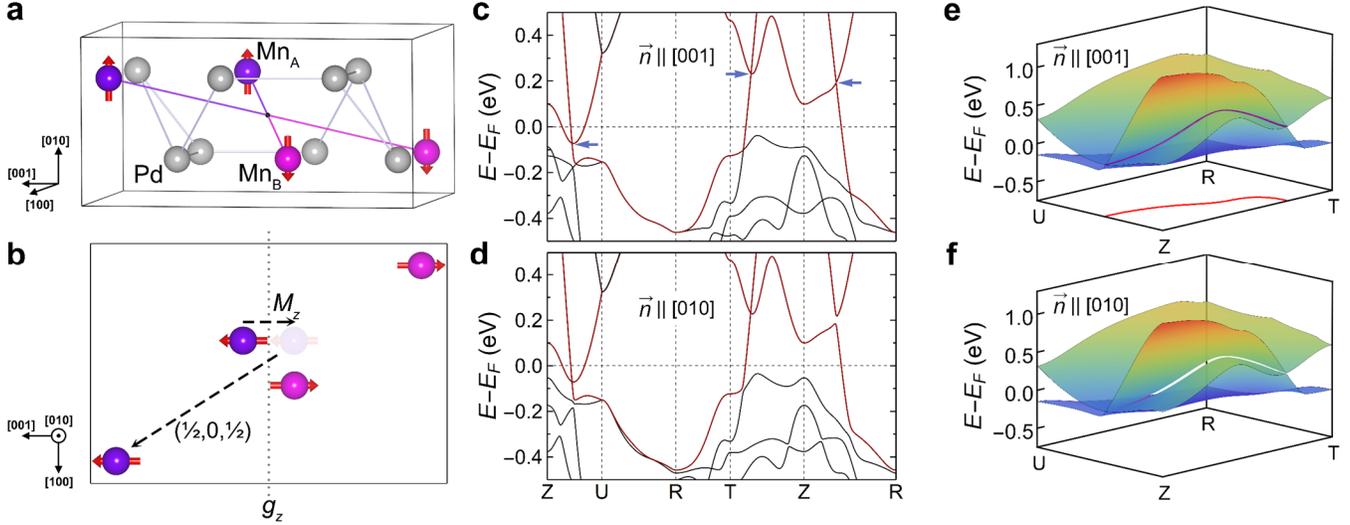

FIG. 2. Néel vector controlled Dirac nodal line in MnPd$_2$. (a) Crystal structure of MnPd$_2$. Two magnetic sublattices Mn$_A$ and Mn$_B$ are connected by the *PT* symmetry. Red arrows indicate the Mn magnetic moments. (b) Schematic of the glide $g_z$ symmetry for $\vec{n} \parallel$ [001], which connects two Mn atoms in the same sublattice via mirror reflection $M_z$ followed by translation (½, 0, ½) (indicated by the black dashed lines). (c, d) Band structure of MnPd$_2$ in the $k_z = \frac{\pi}{c}$ plane for $\vec{n} \parallel$ [001] (c) and $\vec{n} \parallel$ [010] (d). The two bands forming the Dirac nodal line are indicated by dark red color. Arrows in (c) indicate the symmetry protected Dirac band crossings. (e, f) Energy dispersions of the two crossing bands in the $k_z = \frac{\pi}{c}$ plane for $\vec{n} \parallel$ [001] (e) and $\vec{n} \parallel$ [010] (f). The Dirac nodal line and its projection to the $k_z = \frac{\pi}{c}$ plane are shown by the purple and red lines, respectively.

The spin Hall conductivity is given by [35]

$$\sigma_{ij}^k = \frac{e^2}{\hbar} \int \frac{d^3\vec{k}}{(2\pi)^3} \sum_n f_{n\vec{k}} \Omega_{n,ij}^k(\vec{k}), \quad (1)$$

$$\Omega_{n,ij}^k(\vec{k}) = -2 Im \sum_{n' \neq n} \frac{\langle n\vec{k}|J_i^k|n'\vec{k}\rangle \langle n'\vec{k}|v_j|n\vec{k}\rangle}{(E_{n\vec{k}} - E_{n'\vec{k}})^2}, \quad (2)$$

where $f_{n\vec{k}}$ is the Fermi-Dirac distribution function for band $n$ and wave vector $\vec{k}$, $\Omega_{n,ij}^k(\vec{k})$ is the spin Berry curvature, $J_i^k = \frac{1}{2}\{v_i, s_k\}$ is the spin-current operator, $v_i$ and $s_k$ are velocity and spin operators, respectively, and $i, j, k = x, y, z$.

The calculated spin Hall conductivities of MnPd$_2$ for $\vec{n} \parallel$ [001] and $\vec{n} \parallel$ [010] are given in Table 1. Overall, we find that the predicted magnitude of the spin Hall conductivity in MnPd$_2$ is comparable to that in Ta [36,37], a widely used spin current source in spin-orbit torque devices [38], but somewhat smaller than that predicted for Pt [39]. Here, we focus on $\sigma_{xy}^z$ as a representative component of the spin Hall conductivity. The other components are discussed in Supplemental Material [34].

As follows from Eqs. (1,2), the spin Hall conductivity is strongly affected by band anticrossings, where the spin Berry curvature is significantly enhanced when the energy separations between bands $n$ and $n'$ at a given $\vec{k}$ point are small. It is expected, therefore, that switching between the degenerate and gapped states of the Dirac band can lead to a notable change in the spin Hall conductivity, even if such a switchable Dirac point or a Dirac nodal line is buried behind trivial Fermi surfaces [40,41].

TABLE 1. Calculated spin Hall conductivities $\sigma_{ij}^k$ (in units $\Omega^{-1}$ cm$^{-1}$) for two different orientations of the Néel vector in orthorhombic MnPd$_2$: $\vec{n} \parallel$ [001] and $\vec{n} \parallel$ [010].

| $\vec{n}$ | $\sigma_{xy}^z$ | $\sigma_{yx}^z$ | $\sigma_{xz}^y$ | $\sigma_{zx}^y$ | $\sigma_{yz}^x$ | $\sigma_{zy}^x$ |
|---|---|---|---|---|---|---|
| [001] | 155.9 | -170.7 | -104.4 | 175.2 | 120.6 | -66.7 |
| [010] | 232.0 | -176.0 | -134.9 | 138.5 | 112.7 | -66.0 |

In Figs. 3(a,b), we compare the calculated spin Berry curvature $\Omega_{xy}^z$ in the $k_z = \frac{\pi}{c}$ plane at $E = E_F$ for $\vec{n} \parallel$ [001] and $\vec{n} \parallel$ [010]. We find that while $\Omega_{xy}^z$ is small for $\vec{n} \parallel$ [001] in the whole $k_z = \frac{\pi}{c}$ plane (Fig. 3(a)), it exhibits a notable peak for $\vec{n} \parallel$ [010] (Fig. 3(b)). Such a sharp peak appears at different energies and is associated with the Dirac nodal line. As seen from Fig. 3(c), at the points where the nodal line crosses an equi-energy plane the spin Berry curvature is strongly enhanced (as is indicated by the color contrast in Fig. 3(c)). Such a sizable change in the spin Berry curvature affects the spin Hall conductivity. As is evident from Fig. 3(d), within the energy window corresponding to the Dirac nodal line, $\sigma^z_{xy}$ is much larger for $\vec{n} \parallel [001]$ than for $\vec{n} \parallel [010]$, which is due to the gap opening along the dispersive Dirac nodal line as the result of the Néel vector reorientation.

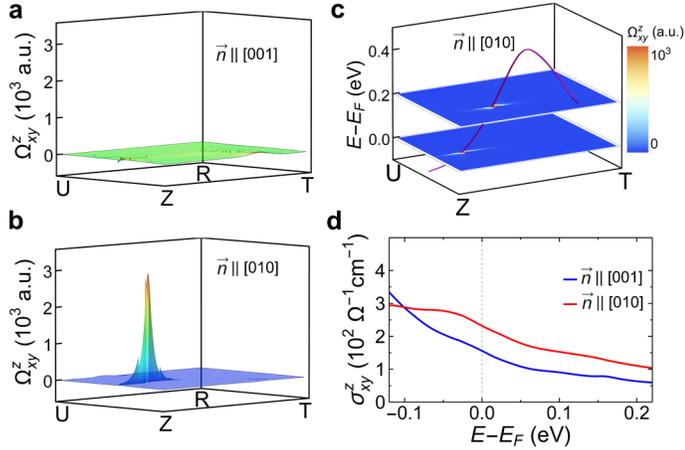

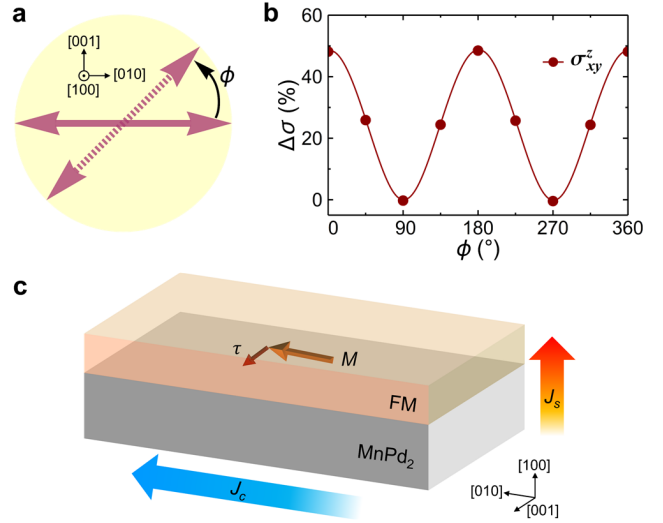

FIG. 3. Néel vector control of the spin Hall effect. (a, b) Calculated spin Berry curvature $\Omega^z_{xy}$ in the $k_z = \frac{\pi}{c}$ plane at $E = E_F$ for $\vec{n} \parallel [001]$ (a) and $\vec{n} \parallel [010]$ (b). (c) The color maps of $\Omega^z_{xy}$ in the $k_z = \frac{\pi}{c}$ planes at two different energies $E = E_F$ and $E = E_F + 0.2$ eV for $\vec{n} \parallel [010]$. Solid purple line represents the nodal line. (d) Spin Hall conductivity $\sigma^z_{xy}$ as a function of energy for $\vec{n} \parallel [001]$ (blue line) and $\vec{n} \parallel [010]$ (red line).

FIG. 4. Angular dependent spin Hall effect and its detection. (a) Reorientation of the Néel vector with angle $\phi$ in the (100) plane relative to its equilibrium [010] direction. (b) Spin Hall conductivity $\sigma^z_{xy}$ as a function of angle $\phi$. Solid line is guide for eye. (c) Schematic of the spin-orbit torque device representing a ferromagnetic layer deposited on $MnPd_2$. The Néel vector in $MnPd_2$ is controlled by charge current $J_c$ producing the Néel spin-orbit torque. Spin Hall current $J_s$ driven by $J_c$ generates torque $\tau$ on magnetization $M$ of the ferromagnet which depends on the Néel vector orientation.

Rotation of the Néel vector in the (100) plane of $MnPd_2$ changes the spin Hall conductivity $\sigma^z_{xy}$ in an oscillatory fashion (Figs. 4(a,b)). The predicted variation of $\sigma^z_{xy}$, reaching the maximum value of about 50%, can be used to experimentally detect the effect of the Néel spin-orbit torque on the Néel vector in $MnPd_2$, as discussed below.

We note that using anisotropic magnetoresistance [6] to detect the effect is problematic due to the uniaxial magnetocrystalline anisotropy of $MnPd_2$ causing the $\vec{n} \parallel [010]$ state to have lower energy than the $\vec{n} \parallel [001]$ state. As a result, affecting the Néel vector can only be produced under conditions of a large steady charge current ($\sim 10^8$-$10^9$ A/cm$^2$) generating the Néel spin-orbit torque. Switching off the charge current leads to the relaxation of the Néel vector back to the equilibrium $\vec{n} \parallel [010]$ direction. This is different from tetragonal CuMnAs [12], which exhibits the bi-axial anisotropy, and thus the Néel vector remains stable after its 90° rotation by the Néel spin-orbit torque and then turning off the charge current.

The spin Hall conductivity under the influence of the Néel vector reorientation can be measured using a spin-orbit torque device shown in Fig. 4c. Here a ferromagnetic layer is deposited on top of the $MnPd_2$ (100) surface, forming a $MnPd_2$/ferromagnet bilayer. Charge current $J_c$ along the [010] direction is driven by an external source to reorient the Néel vector from the easy [010] axis towards the [001] direction. At the same time, the charge current $J_c$ generates spin Hall current $J_s$ flowing in the [100] direction and carrying a spin polarization along the [001] direction. This spin current has conductivity $\sigma^z_{xy}$ which depends of the orientation of the Néel vector, according to Fig. 4(b). The spin current $J_s$ enters the ferromagnetic layer and exerts spin-Hall torque $\tau$ on magnetization $M$ of the ferromagnetic layer. When the charge current density is small, the spin Hall conductivity is constant corresponding to $\phi = 0$. When the current density becomes sufficiently large, the Néel spin-orbit torque reorients the Néel vector away from its equilibrium [010] direction and the spin Hall conductivity $\sigma^z_{xy}$ decreases. Since the sizable change of $\sigma^z_{xy}$ can be obtained even with a small tilting of the Néel vector (Fig. 4b), a moderate charge current is sufficient to confirm our prediction. This variation in the spin Hall conductivity can be detected by various standard techniques such as the spin-torque ferromagnetic resonance (ST-FMR) [42], the magneto-optical Kerr effect (MOKE) [43], and the second-harmonic Hall effect [44].

In a similar way, the Néel vector control of the other components of the spin Hall conductivity listed in Table 1 can be measured using the appropriate design of the spin-orbit torque device. The proposed approach represents a new way to electrically control the spin Hall conductivity *in-situ*. It is different from the recent approaches to tune the spin Hall effect by changing the scattering center density [45,46], varying the chemical composition [47], or ionic gating [48].

The predicted strong dependence of the spin Hall conductivity on the Néel vector is expected to have non-trivial implications for the dynamics of topological textures, such as domain walls or skyrmions [49,50]. This dependence can also be reflected in the dynamics of AFM domain walls under moderate spin-orbit torques [51]. Reversible switching of the Néel vector may be realized using ferroelastic strain from a piezoelectric substrate [52].

Overall, we have demonstrated that $MnPd_2$ is a promising material candidate for topological antiferromagnetic spintronics. On one hand, this material has the required magnetic group symmetry to support the Néel spin-orbit torque, which allows the reorientation of the Néel vector. On the other hand, $MnPd_2$ holds a dispersive Dirac nodal line across the Fermi energy. Gap opening and closing along the Dirac nodal line is controlled by the orientation of the Néel vector, which changes the spin Hall conductivity. The latter can be detected by standard techniques using the proposed spin-orbit torque device. We hope that our theoretical predictions will stimulate the experimentalists to explore the unique properties of $MnPd_2$.

**Acknowledgments** We thank Tianxiang Nan and Mark Rzchowski for helpful discussions. This work was supported by the National Science Foundation (NSF) through the Nebraska MRSEC program (grant DMR-1420645) and the DMREF program (grant DMR-1629270). S.-H.Z thanks the support of National Science Foundation of China (NSFC Grants No. 11504018). Computations were performed at the University of Nebraska Holland Computing Center.

* dfshao@unl.edu
† tsymbal@unl.edu